What? Who? Why? Stellify

J.C. Holbrook
Science, Technology & Innovation Studies
University of Edinburgh

**Abstract:**
In his 1981 article, Roberts highlights the term 'stellify' defined as "to transform (a person or thing) into a star or constellation, to place among the stars." Using the case of the Tabwa people of central Africa, now the Democratic Republic of Congo, Roberts presents among other things the sky as a mnemonic for remembering migrations and remembering culture heroes. We do not know the details of the processes of stellification, however we do know what has been stellified in many cultures by examining their names for stars and asterisms and their skylore. Of the many ideas presented in his latest book, Aveni teases out the ideas of the sky stories having connections to celestial motion, as well as being a mnemonic for remembering seasonal activities and a mnemonic for remembering locally embedded moral, ethical, and sociocultural codes, thus overlapping with Roberts' supposition of the sky serving as a mnemonic. I draw on case studies to flesh out three themes 1. celestial motions, 2. moral, ethical, and sociocultural codes, and 3. seasonal activities within African sky stories. As previously stated, though the human process of assigning names and stories to the night sky as well as stellifying aspects of their lives is not fully understood, these three themes hold promise for being foundational if not part of every culture's practice of stellification.

Keywords: Indigenous Astronomy, Africa, Pleiades, Arcturus, Celestial motions, Morals

Introduction:
Allen F. Roberts provides a definition of 'stellify' in his 1981 article(Roberts 1981) as "to transform (a person or thing) into a star or constellation, to place among the stars." Roberts studied the Tabwe people during the 1970s in central Africa in the country known as Zaire at the time but is now the Democratic Republic of Congo (DRC). Roberts findings about the Tabwa indigenous astronomy are both complicated and multi-layered. He was able to compare his findings to those of missionaries nearly a century earlier, thus allowing for consideration of changes over time. By way of introduction, two celestial bodies as examples of stellification are Orion's Belt and the Milky Way. The recorded names for Orion's Belt in the traditions of the Tabwa are Luziwe, Luzigwe and Kabwe. As with many African cultures, Orion is associated with hunting (Vieira 2009; d'Huy 2013), for the Tabwa the stars of the belt (east to west) are a hunter, a dog, and a bush rat. The hunter and dog chase the bush rat that is fleeing to the west. Kabwe means dog and the term refers to the entire asterism of Orion's belt; it was the name used during Roberts' fieldwork in the 1970s. The older names of Luziwe and Luzigwe, refers to a path but also may refer to food and other restrictions that the chief had to follow while the Tabwa were migrating from the north to their present location. Another cultural hero is the

Aardvark. In their origin myth, the first being was the Aardvark with its hunting dogs. The tunnelling Aardvark arrived in a beautiful new land but was lonely. The creator deity gave Aardvark human companionship in the form of a man and woman along with a basket that contained the Sun, Moon and stars. The Aardvark may be the hunter star in the belt of Orion, but the name for the Milky Way is Mulalambo which means Aardvark's tunnel. Among Roberts' conclusions is that cultural heroes such as the Aardvark and the chief that lead their migration are found in the stars of Orion's belt and the Milky Way, both found within the layers of meaning associated with the idea of paths, journeys, and paths in the sky.

> *"Ptolemy, a second-century Alexandrian astronomer, listed forty-eight constellations. Nearly three dozen were named after land animals, fish, and birds, with a sprinkling of serpents and humanoids—as well as one insect."* (Aveni 2019, 3)

Cross cultural comparisons of constellations present much to think about in terms of what humans find important enough to stellify. Anthony F. Aveni in his latest book declares that the most numerous beings that are constellations in Ptolemy's sky are animals, fish and birds (Aveni 2019). This sky filled with the natural world did have a few humans: Andromeda – the princess in the Perseus myth, Aquarius – the water bearer from the Ganymede myth, Auriga – the Charioteer, Boötes – the Herdsman, Cepheus – the King in the Perseus myth, Cassiopeia – the Queen in the Perseus myth, Gemini – the brothers Castor (made immortal by Zeus) and Polydeuces (fathered by Zeus), Hercules – culture hero, Ophiuchus – the Serpent bearer, Perseus – culture hero. Ten out of forty-eight constellations are or were humans, with four of these connected to the myth of Perseus (there is also Perseus' magical steed, Pegasus, and the sea monster that threatened Cassiopeia, Cetus, also as constellations). Reflections of myths, heroic people and the natural environment, this paints with a very broad brush some of the things that people turn into constellations. However, Aveni shows how specific sky myths and their associated constellations serve three deeper functions: 1) they encode moral and ethical behaviour, 2) they reflect celestial motions, and 3) they are connected to seasonal activities. Using the example of Orion, which is a constellation that is recognizable due to its many bright stars and it can be seen from both the Northern and Southern Hemispheres so has legends from both, Aveni explains how the constellation exiting from the night sky at the end of Northern Hemisphere winter and returning in the Northern Hemisphere autumn is captured in the sky myths, how the exit and return are connected to fertility and agricultural cycles, and how the exit is often a punishment and the return a redemption moment in the sky myths. Returning to the Tabwe example of Orion's Belt, in addition to stellifying heroes and their journeys, it was speculated that part of the moral and ethical behaviour of chiefs to follow the food and other restrictions making them worthy of being chiefs. The orientation of the Milky Way in the evening signalled to the Tabwe people the change of seasons, i.e. a connection to seasonal activities.

Focusing on the African continent, this article is an exploration of the three themes of stellification that emerged through reading Roberts and Aveni, with the caveat that these are not all of the possible themes presented in either work: 1. celestial motions, 2. moral, ethical, and sociocultural codes, and 3. seasonal activities within African sky stories.

Celestial Motions:
The Luba (Baluba) people of the Democratic Republic of Congo, have the story of why humans die, in short because they are liars. Their creator God, Fidi Mukullu, sent the Sun to fetch him palm wine, the Moon to fetch him palm wine, the Pleiades to fetch him palm wine, and a human to fetch him palm wine. Upon their return, Fidi Mukullu placed the Sun, Moon, Pleiades and the human in a special ditch in which liars could not emerge and asked them if they had drunk the palm wine. The Sun emerged and as a result can be reborn daily. The Moon emerged and as a result is reborn monthly. The Pleiades emerged and as a result is reborn every rainy season. Only the human did not emerge, thus Fidi Mukullu declared that humans must die (Klipple 1938, 754).

The Fidi Mukullu myth presents Luba knowledge of celestial motions starting with the Sun. If the special ditch where Fidi Mukullu placed everyone is equivalent to an absence from the visible sky, then being reborn is a return to the sky. The order in which the celestial bodies emerged from the ditch is reflective of how long they are absent from the sky. The Sun returns daily, the moon rises and sets but isn't altogether absent except during the new moon phase in which it is not visible for a few days. The Pleiades star cluster, like the constellation Orion, is visible from both the Northern and Southern Hemispheres. It is a bright asterism that is easily recognizable in dark sky locations. The return of a star or asterism to the night sky is called the Heliacal rise, the star or asterism rises on the eastern horizon just before the Sun rises. For 2021, the Heliacal rise of the Pleiades is June 15th, the Pleiades are visible in the night sky on the eastern horizon at sunset in mid-November and the estimated date for exiting the sky, i.e. no longer visible on the western horizon after sunset, is in mid-April.

The Tamberma people of Togo practice a pantheistic religion with many of their deities having a connection to a particular star or asterism (Blier c1986). The Tamberma deities and their celestial behaviours determine their agricultural and ceremonial activities. When a star or asterism connected to a particular deity disappears from the night sky at sunset on the western horizon, that deity begins its walk on the surface of the Earth from west back to the east. This time is considered the ceremonial season for that deity. When the deity reaches the east and reappears in the sky, this is the culmination of their ceremonial season. One star that is rarely mentioned in African Indigenous Astronomy is Arcturus. Arcturus is the brightest star in the northern constellation of Boötes and it is the fourth brightest star in the night sky after Sirius, Canopus and Alpha Centauri. The Tamberma God of war and death is Fayenfé and is associated with Arcturus. When Fayenfé/Arcturus leaves the sky, it is time to harvest yams and for the first fruits ceremonies as well as traditionally a time of war and violence. Fayenfé while walking the Earth was known to possess people pushing them to enact random acts of violence, because of this people seldom left their villages during this time. The celestial motions encoded into the Tamberma religion are the annual heliacal risings and settings of stars and asterisms.

The Bamana people of Mali have a culture that is structured by six initiation societies. The Tyiwara society is focused on agriculture and working in the fields, but also conveys the relationship between the Sun and 'the nurturing Earth' (Zahan 2005). A mythical antelope figure is the tyiwara/ciwara/chiwara, the excellent farmer that taught the Bamana people agricultural cultivation (Wooten and Roberts 2000, 31). The Tyiwara is artistically rendered in two headdresses that represent the Sun and the Earth. The Tyiwara dance, where these headdresses are part of the costumes worn, happens as people are preparing their fields for planting. The Sun headdress is the male Tyiwara and the Earth headdress is the female. The headdresses are works of art in themselves (see Figure ?), the male headdress has a mane that incorporates a zigzag pattern. This has been interpreted as capturing the annual north – south – north motion of the Sun on the horizon, which is further reinforced in the Tyiwara performance in which the Sun dancer moves in a zigzag pattern. The zigzag pattern is how an antelope runs away as well. This zigzag pattern is called *tlé sirá* (the march of the sun) (Zahan 1980, 91). The female Earth headdress is often depicted as more horizontal than vertical with a smaller baby antelope on its back. The celestial motion captured by the Tyiwara ritual performance is the apparent motion of the Sun relative to the Earth in the zigzag patterns danced and the zigzag pattern on the Sun/male headdress.

Cultural Moral Codes and Correct Behavior:
A Khoikhoi legend from southern Africa relays the story of the hopeless hunter (Snyman n.d.) also called the Curse of the Women (Hahn 1881, 74). The hunter was the husband, aob (the star Aldebaran), in a polygamous marriage with a group of women, ǀKhunuseti (the Pleiades). They pushed him to go hunting to bring home meat. He set out after a herd of zebra, !goregu (Orion's Belt), and shot his one arrow, ≠ab (iota, theta, and 42 Orionis (Alcock 2014, 288)), and missed. The arrow landed close to where a lion, xami (Betelgeuse), was hunting the same herd of zebra. The hunter couldn't retrieve his arrow because of the lion, so he remains in the sky waiting. His wives had warned him not to return home empty handed. They said, "Ye men, do you think that you can compare yourselves to us, and be our equals? There now, we defy our own husband to come home because he has not killed game (Hahn 1881, 74." The hopeless hunter myth encodes the need for men to support their family by supplying fresh meat through being a good hunter. Though the KhoiKhoi associate the Pleiades with their rainy season and other seasonal activities, that knowledge is not part of this particular myth.

Returning to the Fidi Mukullu story of the Luba people, telling the truth is a moral code that is the fabric of the story. Those that tell the truth are placed in the sky with the Pleiades being stellified. The human that lies is punished with having to die. The Tyiwara society places emphasis on agriculture and its importance to the Bamana people as reflected in their Sun and Earth, male and female, elements. Turning to the Tamberma people, Fayenfé/Arcturus is a violent God of war and death. During the time Fayenfé is walking the Earth, people have to be careful to not become violent. In fact, they enact several rituals including making offerings to Fayenfé to appease Fayenfé's need for war and violence. This points to a cultural code of non-violence.

Seasonal Activities:
Much of the information available on seasonal activities in Africa and the sky is focused on predicting the weather, in particular predicting rain or the start of the rainy season, in connection with agricultural activities. The Pleiades' name stellifies this concept in many African cultures. For example, in the traditions of the South African Zulu people the Pleiades is iSilimela (the digging stars), for the Sotho and Tswana peoples of South Africa the Pleiades is selemela (cultivator), in Swahili, the trade language along the eastern coast of Africa for centuries, the Pleiades is Kilimia (the digging stars). It is thought that these related names comes from a common source that was carried into central, eastern and southern Africa with the migration of Bantu peoples (Snedegar 1995, 533). Though the agricultural seasons for these locations are different, local peoples adjusted the times when the Pleiades were observed and the correct position of the Pleiades in the sky to match their agricultural season. For example, the Pleiades have to be overhead at sunset or the Pleiades have to be visible in the east at sunset. In this way, the Digging stars stellification remained accurate.

Returning again to the Fidi Mukullu story of the Luba people, focusing on what is said about the Pleiades in particular, the Pleiades is connected to the rainy season. Interestingly, since DRC straddles the equator, the north of the country has a dry season that is from November to March, and the south, where the Luba live, is reversed so the rainy season is November to March. It can be concluded that for the Luba, the November appearance of the Pleiades at sunset is what is used for indicating the start of the rainy season. Therefore, its evening appearance marks the end of the dry seasons' activities and the beginning of the rainy season activities. For the Tamberma people, their deities with associated stars/asterisms as well as their celestial motions determine seasonal activities such as those described for Fayenfé/Arcturus, which include harvesting yams and first fruits celebrations and ceremonies. The Bamana people, through their Tyiwara society, celebrate the time to begin their annual cultivation.

Stellification:
There are many myths associated with the constellation Boötes, the Greek myth has to do with Zeus's son with Callisto, Arcas. Arcas became a King and the region Arcadia in Greece was named after him. To save Callisto from Hera's, Zeus' wife, rage, Zeus changed Callisto into a bear. King Arcas came across the bear and mistaking it for a normal bear attempted to kill it. In order to save both Callisto and Arcas, Zeus changed both into constellations: Callisto into the great bear (Ursa Major) and Arcas into Boötes with Arcturus (the guardian of the bear) as the brightest star. The stellification of Callisto and Arcas was to protect them but also reflects the remorse that Zeus felt for his actions. Arcturus for the Tamberma people is a war god and bringer of death. When Arcturus is in the sky, the Tamberma people are safe, but when absent from the night sky, the war god is walking the Earth and people need to beware. Arcturus/Fayenté is stellified as a deity but also is a warning to the Tamberma people, with the additional function of marking when yams are to be harvested.

Stellification as explored through the three themes of celestial motions, moral codes and seasonal activities finds purchase in the skylore of Africans. From the African cultures explored in this article, celestial motions such as the daily motion of the Sun, the annual motion of the Sun, the monthly motion and phases of the Moon, the rising and setting of stars and asterism, and the shifting of the Milky Way entertainingly appear in the African legends and myths presented. Gods, cultural heroes, and animals are stellified to remind people of correct behavior including warning people of bad behaviors such as being an unsuccessful hunter in the Khoikhoi myth. The agricultural cycle which is connected to rainfall is important for survival in Africa, thus many of the seasonal activities aided by the sky involve sighting familiar stars and asterisms, especially the Pleiades, that coincide with the beginning of the rainy seasons. These three themes within stellification are neither exclusive nor exhaustive, other details of stellification could be to tease out if people are stellified only as a reward or only for punishment in certain cultures, then there is stellification as a means of escape. Given the plethora of animals that are stellified, what makes certain animals worthy of stellification especially since sky animals often include both wild and domesticated animals? These questions and themes of stellification should be considered during the continued study of African Indigenous astronomy.